\documentclass{ws-procs9x6}

\begin{document}

\title{Preheating and Thermalization after Inflation%
% \footnote{\uppercase{T}alk presented at the
% \uppercase{W}orkshop on \uppercase{S}trong and 
% \uppercase{E}lectroweak \uppercase{M}atter
% (\uppercase{SEWM 2002}), \uppercase{O}ctober 2-5, 2002, 
% \uppercase{H}eidelberg, \uppercase{G}ermany.}
}

\author{Raphael Micha}

\address
    {%
    Theoretische Physik, 
    ETH Z\"urich, 
    CH-8093 Z\"urich, 
    Switzerland 
    }%

\author{Igor I. Tkachev}

\address 
    {%
      Theory Division, CERN, CH-1211 Geneva 23, Switzerland \\
      and \\
      Institute for Nuclear Research of the Russian Academy of
      Sciences, 117312, Moscow, Russia
    }%


\maketitle

\abstracts {% 
After a short review of inlationary preheating, we discuss the
development of equilibrium in the frameworks of massless
$\lambda \Phi^4$ model. It is shown that the process
is characterised by the appearance of Kolmogorov spectra and the
evolution towards thermal equilibrium follows self-similar dynamics.
Simplified kinetic theory gives values for all characteristic
exponents which are close to what is observed in lattice simulations.
This allows estimation of the resulting reheating temperature.
}%

\section{Introduction}

The dynamics of equilibration and thermalization of field theories is
of interest for various reasons.  In high-energy physics understanding
of these processes is crucial for applications to heavy ion collisions
and to reheating of the early universe after inflation (for review of
inflation see\cite{books}). Inflationary
problems for which this understanding can be crutial include
\begin{itemize} 
\item Baryogenesis. Indeed, generation of baryon
  asymmetry is possible only in a state which is out of equilibrium.
  Universe is in highly non-equilibrium state after
  preheating. This opens up new possibilities and mechanisms for
  baryogenesis\cite{baryo}. Clearly, the final unswer should
  depend on details of the equilibration process in many ``new'' and
  ``old'' (e.g. Affleck-Dine type\cite{Affleck:1984fy}) models.
\item The problem of over-abundant gravitino production in
  supergravity models\cite{gravitino} is avoided if reheating 
  temperature is sufficiently low.
\item The abundance of superheavy dark matter and other relics
  depends on the time of transition from matter to radiation dominated
  expansion during reheating of the universe\cite{SHDM}.
\end{itemize} 
Thermalization of field theories was discussed
in Refs.\cite{thermalization}.  However, at present
the process of thermalization after preheating is still far away from
being well understood and developed. The problem is that at the
preheating stage the occupation numbers are very large, of order of
the inverse coupling constant. In addition, in many models the coherent
inflaton oscillations does not decay for a long time. Therefore, 
a simple kinetic approach is not applicable.

Fortunately, the description in terms of classical field theory is
valid in this situation\cite{Khlebnikov:1996mc}, and the process of
preheating, as well as subsequent thermalization, can be studied on a
lattice.  In the papers\cite{Felder:2000hr,Micha:2002ey} this
approach was adopted. The goal is to integrate the system on a lattice
sufficiently accurately and sufficiently far in time to be able to see
generic features, and possibly to the stage, at which the kinetic
description becomes a good approximation scheme. Several generic rules
of thermalization were formulated in Ref.\cite{Felder:2000hr}, like
the early equipartition of energy between coupled fields.  However,
the problem is very complicated and there are other unanswered
important questions, including: what is the final thermalization
temperature, at what stage the kinetic description becomes valid, what
is the functional form of particle distributions during the
thermalization stage, etc. These issues were addressed in
Ref.\cite{Micha:2002ey} in frameworks of the ``minimal'' inflationary
model, the massless $\lambda \Phi^4$-theory.

It was shown that the particle distribution function follows a {\em
self-similar} evolution related to the turbulent transport of wave
energy. This allows estimation of the reheating temperature, which
turns out to be very low in the consiered model.  The appearence of
self-similar regime of evolution should be generic for a wide class of
models since typical ranges of particle momenta at preheating and in
thermal equilibrium are widely separated. 

In the first part of this talk a short review of preheating after
inflation is given. In the second part we report on the recent progress in
understanding of thermaliztion after preheating.

\section{Preheating}

Inflation solves the flatness and horizon problems of the standard big
bang cosmology and provides a calculable mechanism for the generation
of initial density perturbations\cite{books}.  At the end of inflation
the Universe was in a vacuum-like state.  In the process of decay of
this state and subsequent thermalization (reheating) all matter
content of the universe is created. It was realized recently that the
initial stage of reheating, dubbed preheating\cite{preheat}, is a very
fast process. This initial stage by now is well 
understood.\cite{preheat,Kofman:1997yn,Khlebnikov:1996mc}$^{,12-17}$

%Prokopec:1996rr,hybrid,fermions1,Giudice:1999fb}.

\subsection{Fast Inflaton Decay}

Bose-stimulation aids the process of creation of bosons.  Occupation
numbers grow exponentially with time, $n = e^{\mu t}$, which results
in a fast, explosive decay of the inflaton\cite{preheat,Kofman:1997yn} 
and creates large classical
fluctuations at low momenta for all coupled Bose-fields. This can have
a number of observable consequences. Examples include non-thermal
phase transitions\cite{non-termal-pt}, generation of a
stochastic background of the gravitational
waves\cite{Khlebnikov:1997di}, 
and a possibility for a novel mechanism of
baryogenesis\cite{baryo}. The problem is complicated, but
fortunately, the system in this regime of particle creation became
classical and can be studied on a 
lattice\cite{Khlebnikov:1996mc,Prokopec:1996rr,Felder:2000hq}.

On the other hand, occupation numbers of fermions satisfy $n \le 1$ at
all times because of the Pauli blocking. This may create an impression
that the fermionic channel of inflaton decay is not important. This is
a false impression\cite{fermions1}.  Production of supreheavy
fermions can be more efficient compared to
bosons\cite{Giudice:1999fb}. Let us explain why and when this happens.

\subsection{Fermions versus Bosons}

The effective mass of a scalar $X$ (interaction Lagrangian $~L_{\rm
int}=\frac{1}{2} g^2 \phi^2 X^2$) and of a fermion $\psi$ (interaction
Lagrangian $~L_{\rm int} = g \phi \bar{\psi}\psi$) coupled to the
inflaton field $\phi$ are given by the following expressions
\begin{eqnarray}
{\rm scalar~} X:&& \;\; m^2_{\rm eff} = m_X^2 + g^2\phi^2(t)\,  
\label{mX} \\
{\rm ~fermion~} \psi:&& \;\; m_{\rm eff} = m_\psi + g\phi(t)\, . 
\label{mF}
\end{eqnarray}
The effective mass of a scalar $X$ depends quadratically upon the inflaton
field strength and therefore effective mass is always larger than the
bare mass $m_X$. In the case of fermions the inflaton field strength
enters linearly and the effective mass can cross zero. Therefore,
superheavy fermions can be created during these moments of zero
crossing\cite{Giudice:1999fb}. Indeed, it is easier to create a light
field and it is the effective mass which counts at creation.

The Coupling $g$ by itself is not relevant for the process of creation,
$g$ always comes in combination with the inflaton field strength. To make
dimensionless relevant combination we have to rescale $g\phi$ by a
typical time scale of creation.  In the present case this
will be the period of inflaton oscillations or inverse inflaton mass.
We obtain
\begin{equation} 
g^2 \rightarrow  q \equiv \frac{g^2\phi^2}{4m_\phi^2}.
\end{equation}
The parameter q determines the strength of particle production caused by
the oscillations of the inflaton field. It can be very large even when
$g$ is small since ${\phi^2}/{m_\phi^2} \approx 10^{12}$.

\begin{figure}[ht]
\centerline{\epsfxsize=3.9in\epsfbox{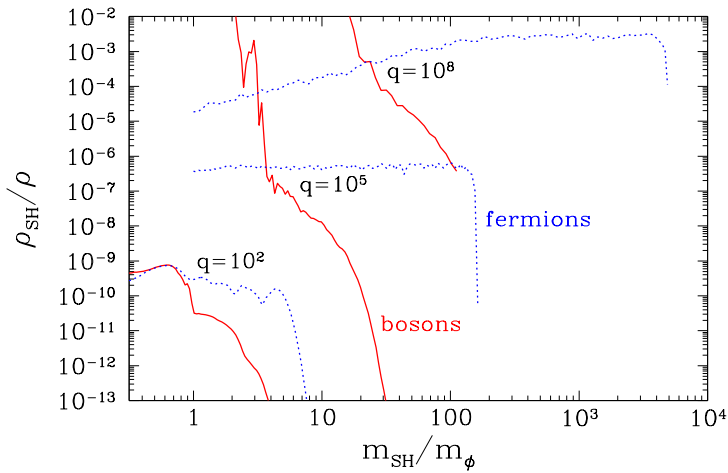}}   
\caption{Effectiveness of Super-Heavy particle production.
Solid lines production of fermions, dotted 
lines production of bosons.
\label{fig:BvF}}
\end{figure}

A fraction of the initial energy density which goes respectively to
bosons and fermions is plotted in Fig.~\ref{fig:BvF} for several
values of $q$. For fixed $q$ this fraction is a function of the ratio
of the particle mass, $m_{\rm SH}$, to the inflaton mass. We see that
superheavy fermions are more efficiently created compared to bosons,
indeed.

%%%%%%%%%%%%%%%%%%%%%%%%%%%%%%%%%%%%%%%%%%%%%%%%%%%%%%%%%%%%%%%%%%%%%%%%%
%%%%%%%%%%%%%%%%%%%%%%%%%%%%%%%%%%%%%%%%%%%%%%%%%%%%%%%%%%%%%%%%%%%%%%%%%%

On the other hand, bosons of moderate or small mass (compared to
inflaton), and sufficiently large coupling ($q > 10^4$, which
corresponds to $g^2 > 10^{-8}$) are produced explosively. Their
density grows exponentially with time until back reaction becomes
important.  Corresponding field variances, $\langle X^2 \rangle$, can
reach large values, larger than in a thermal equilibrium at the same
energy density leading to interesting physical consequenses which were
described above.

\section{Thermalization}

A study of thermalization requires long and accurate integration on a
reasonably large lattice. Therefore we consider the simplest model
for reheating after inflation: the massless $\lambda\Phi^4$ theory.

\subsection{The Model}

With conformal coupling to
gravity and after rescaling of the field, $\varphi \equiv \Phi a$,
where $a(t)$ is the cosmological scale factor, the equation of motion
in comoving coordinates describes $\varphi^4$-theory in Minkowski
space-time, $\Box \varphi + \lambda \varphi^3 = 0$.
At the end of inflation the field is homogeneous, $\varphi = \varphi_0
(t)$.  Later on fluctuations develop, but the homogeneous component of
the field which corresponds to the zero momentum in the Fourier
decomposition may be dynamically important and is referred to as the
``zero-mode.''  In such situations it is convenient to make a further
rescaling of the field, $\phi \equiv \varphi/\varphi_0(t_0)$, and of
the space-time coordinates, $x^\mu \rightarrow \sqrt{\lambda}
\varphi_0(t_0) x^\mu$. This transforms the equation of motion
into dimensionless and parameter free form, $\Box \phi + \phi^3 = 0$.
Here $t_0$ corresponds to the initial moment of time, which is the end of
inflation for us here. In what follows we denote dimensionless
time as $\tau$.  With this rescaling the initial condition for the
zero-mode oscillations is $\phi_0(\tau_0 ) = 1 $.  All model
dependence on the coupling constant $\lambda$ and on the initial
amplitude of the field oscillations is encoded now in the initial
conditions for the small (vacuum) fluctuations of the field with
non-zero momenta\cite{Khlebnikov:1996mc}.  The physical normalization
of the inflationary model corresponds to a dimensionful initial
amplitude of $\varphi_0(t_0) \approx 0.3 M_{\rm Pl}$ and a coupling
constant $\lambda \sim 10^{-13}$.

\subsection{Numerical Procedure and Results}

Various quantities were measured on a 3-D cubic lattice and monitored
both in configuration space (zero mode, $\phi_0 \equiv
\langle\phi\rangle$, and the variance, $var(\phi) \equiv
\langle\phi^2\rangle -\phi_0^2$) and in Fourier space, where the wave
amplitudes (which correspond to annihilation operators in the quantum
problem) were defined as
\begin{equation}
a({\vec k}) \equiv \frac{\omega_k \phi_{\vec k} +i \dot{\phi}_{\vec k}}
{{(2\pi)^{3/2}}\sqrt{2\omega_k}} \; .
\end{equation}
The effective frequency $\omega_k \equiv \sqrt{k^2 +m_{eff}^2}$ is
determined by the effective mass
$m_{eff}^2=3\lambda\langle\phi^2\rangle$.  Low order correlators
containing $a_k$, namely $n(k) \equiv \langle a^\dagger a\rangle$,
$\sigma (k) \equiv \langle a a\rangle$ and $\langle a^\dagger a^\dagger a
a\rangle$ were measured.  The first one, which corresponds to the
particle occupation numbers, is of prime interest.

\begin{figure}[ht]
\centerline{\epsfxsize=3.9in\epsfbox{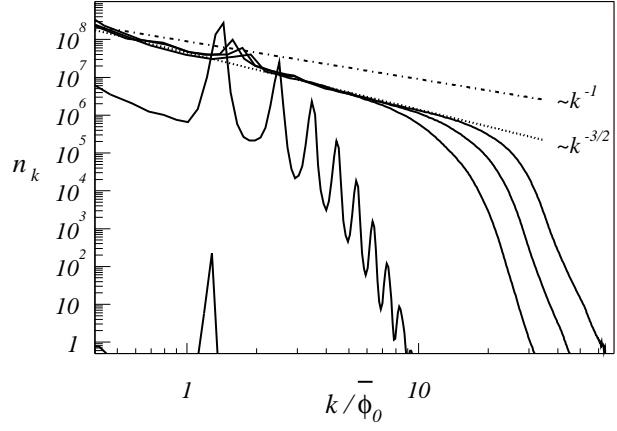}}   
\caption{Occupation numbers as  function
  of $k/\overline\phi_0$ at  $\tau = 100, 400, 2500, 5000, 10000$.
\label{spectra}}
\end{figure}

Occupation numbers at different moments of time are shown in
Fig.~\ref{spectra}. The first stage of the inflaton decay
($\tau<1500$) is characterized by a peaky structure of the
spectra. The first peak, which corresponds to the parametric
resonance, is initially at the theoretically predicted value of $k \sim
1.27$ \cite{Kofman:1997yn}. Higher peaks appear later ($\tau>300$)
because of re-scattering of waves from the first resonant peak. With
time the first peak moves to the left reflecting the change in the
effective frequency of inflaton oscillations.  However, after
rescaling of particle momenta by the current amplitude of the zero
mode oscillations, as it is done in Fig.~\ref{spectra}, the position
of the resonance is approximately unchanged.

\begin{figure}[ht]
\centerline{\epsfxsize=3.9in\epsfbox{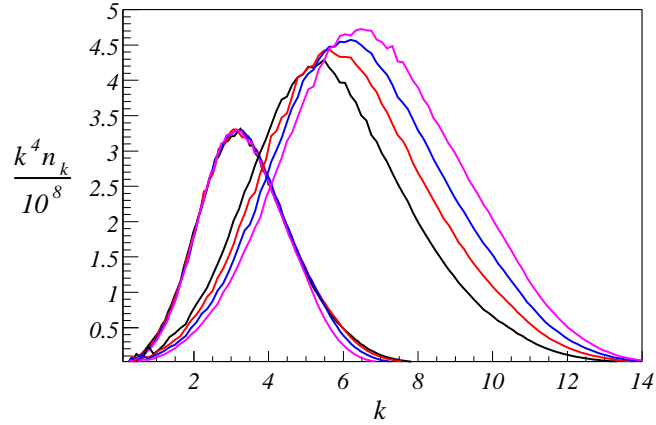}}   
\caption{On the right hand side we plot
    the wave energy per decade found in lattice integration
    at  $\tau = 3600, 5100, 7000, 10000$. On the
    left hand side are the same graphs transformed according to
    the relation inverse to Eq.~(\ref{SelfS}).
\label{self_sim}}
\end{figure}

The second stage of the evolution ($\tau>1500$) has following 
characteristic features\cite{Micha:2002ey} :
\begin{enumerate}
\item The system is statistically close to Gaussian distribution of
  field amplitudes and conjugated momenta, while phases are random,
  see also\cite{thermalization,Felder:2000hr}.
\item In accord with this, the ``anomalous'' correlators are small (but
  non-vanishing), e.g. 
$\sigma_k\sim 10^{-2}\ n_k$.
\item The zero-mode $\phi_0$ and variance $var(\phi)$ are decreasing
according to the power-laws $\phi_0\sim\tau^{-1/3}$ and
$var(\phi)\sim\tau^{-2/5}$.
\item The zero-mode is in a non-trivial dynamical equilibrium with the
bath of highly occupied modes: when the zero-mode is artificially
removed, it is recreated on a short time-scale (Bose condensation).
\item The spectra in the dynamically important region can be described
by a power low, $k^{-s}$ with $s\sim 1.5 - 1.7$.  We see that, despite
the feature (1) holds, the system is not in a thermal
equilibrium. Rather, particle distributions correspond to Kolmogorov
turbulence.
\item The power law is followed by an exponential cut-off at higher $k$.
Energy accumulated in particles is concentrated in the region were the
cut-off starts. It's position is monotously growing, reflecting the
motion towars thermal equilibrium.
\item This motion can be described as a self-similiar evolution
\begin{equation}
n(k,\tau) = \tau^{-q} n_0 (k \tau^{-p}) \, ,
\label{SelfS}
\end{equation}
the best numerical fits are $q \approx 3.5p$ and $p \approx1/5$, see
Fig. \ref{self_sim}.
\end{enumerate} 

%%%%%%%%%%%%%%%%%%%%%%%%%%%%%%%%%%%%%%%%%%%%%%%%%%%%%%%%%%%%%%%%%%%%%%%%%%%%
%%%%%%%%%%%%%%%%%%%%%%%%%%%%%%%%%%%%%%%%%%%%%%%%%%%%%%%%%%%%%%%%%%%%%%%%%%%

%%%%%%%%%%%%%%%%%%%%%%%%%%%%%%%%%%%%%%%%%%%%%%%%%%%%%%%%%%%%%%%%%%%%%%%%%%%%%%
%%%%%%%%%%%%%%%%%%%%%%%%%%%%%%%%%%%%%%%%%%%%%%%%%%%%%%%%%%%%%%%%%%%%%%%%%%%%%%%

\subsection{Discussion}

Let us see if these numerical results can be understood analytically.
Features (1) and (2) facilitate the use of a simple kinetic approach,
$\dot{n_k} = I_k\label{kin1}$, where the collision integral for a
$m$-particle interaction is given by\cite{Micha:2002ey}
\begin{equation}
I_k = \int d\Omega_k\, U_k\, F[n] \, .
\label{kin_coll_Int}
\end{equation}
In $d$ spatial dimensions the integration measure $d\Omega_k$
corresponds to $m-1$ integrations over $d$-dimensional Fourier
space. Included in it is the energy-momentum conservation
$\delta$-functions but not the relativistic $1/\omega (k_i)$
``on-shell'' factors, which instead appear in the ``matrix element''
of the corresponding process, $U_k$. This
convention makes the discussion of relativistic and
non-relativistic cases uniform. The function $F[n]$ is a sum of
products of the type $n_{k_{j}}^{-1}\prod_{i=1}^{m}n_{k_i}$, where $j
\in\{1,\dots,m\}$ with appropriate signs and permutations of indices
for incoming and outgoing particles.  All dynamical aspects of
turbulence follow from the scaling properties of the
system\cite{Turb,Semikoz:1994zp}. Let $\omega_k, n_k$ and $U_k$ have 
defined weights under a $\xi$-rescaling of the Fourier space,
\begin{eqnarray}
\omega(\xi k_i)&=& \xi^{\alpha}\omega (k_i) \; , \nonumber \\
U(\xi k_1, \dots, \xi k_m) &=& \xi^{\beta} U(k_1, \dots, k_m)\; , \nonumber \\
n(\xi k_i) &=& \xi^{\gamma} n(k_i) \; . \label{pow_law} 
\label{scalins}
\end{eqnarray}
The weight of the full collision integral under this reparametrization is
\begin{equation}
I_{\xi k}=\xi^{d(m-2)-\alpha+\beta+(m-1)\gamma }I_{k} \, .
\end{equation}
Then stationary turbulence with constant energy flux over momentum
space is characterised by a power-law distribution function, $n_k\sim
k^{-s}$, where $s=d+\beta/({m-1})$. The scaling properties also give
the exponents of the self-similar distribution, Eq. (\ref{SelfS}).  A
system with energy conservation in particles and $\xi = \tau^{-p}$ has
$q = 4p$ and $p = 1/((m-1)\alpha-\beta)$, while for stationary
turbulence exponent $p$ should be $(m-1)$ times larger.

For a massless $\lambda \phi^4$-theory in three spatial dimensions and
four-particle interaction we have $m=4$, $\beta=-4\alpha$ and $\alpha
= 1$.  In this case $s = 5/3$ and $p = 1/7$ \cite{Micha:2002ey}.
Obtained value for p is somewhat smaller then observed numerically, $p
= 1/5$. For three-particle interaction (the fourth particle belongs to
the condensate in this case and the matrix element contains an
additional factor $\overline\phi_0^2$) the stationary turbulence is
characterized by $s = 3/2$ and $p$ has even smaller value compared to
4-particle interaction.  Numerical results\cite{Micha:2002ey} do not
allow to distinguish between $5/3$ and $3/2$, $s$ rather fluctuates
between these two numbers, while $1/7$ for $p$ gives a fit to the data
not as good as displayed in Fig.~\ref{self_sim}.  However, taking into
account the non-stanionarity of the source (influx of energy from the
zero-mode is still non-negligible) we obtain $q \approx 3.5p$ and $p
\approx 1/6$ \cite{Micha:2002ey}.  This should be considered as
satisfactory agreement between numerical experiment and a theory,
given the simplifications which were made.

\subsection{Equilibration time and temperature}

At late times, with decreased influence of the zero-mode, our simple
analytical description should become more precise.  On the other hand
we can expect that the self-similarity persists. In this case solution
Eq.~(\ref{SelfS}) with $p=1/7$, which determines the transport of
energy to higher momenta, should be valid. In classical frameworks it
will be valid at all later times. Therefore, the time needed to reach
equilibrium can be defined as the time when classical approximation
breaks down. In other words, equilibrium will be reached after
occupation numbers in the region of the peak, see Fig.~\ref{self_sim},
will become of order one.  Using Eq.~\ref{SelfS} we
find\cite{Micha:2002ey} for the equilibration time
$\tau\sim\lambda^{-7/4}\sim 10^{23}$, where in the second equality the
normalization to the inflationary model is assumed.  Rotating back
from the conformal reference frame, we obtain for the reheating
temperature $T_R\sim k_{max}/a(\tau) \sim \lambda^2 \varphi_0(t_0)\sim
10^{-26}M_{\rm Pl}\sim 100$ eV \cite{Micha:2002ey}.
This result coinsides with what could have been obtained in ``naive''
perturbation theory. Namely, equating
the rate of scattering in thermal equilibrium to the Hubble expansion
rate we get in this model $T \sim \lambda^2 M_{\rm Pl}$.  

\section{Conclusions}

Reheating after preheating appears to be a rather slow process.
Although the ``effective temperature'' measured at low momentum modes
during preheating may be high, in the model we have considered the
resulting true temperature is parametrically the same as what could
have been obtained in ``naive'' perturbation theory. We
anticipate this result should be applicable to more realistic models
of inflation. Note that realistic models involve many fields and
interactions and the largest coupling constants will determine the true
temperature.

\section*{Acknowledgments}

R.M. thanks the Tomalla Foundation for financial support.

\end{document}